\documentclass[fleqn,usenatbib]{mnras}

\usepackage{newtxtext,newtxmath}
\usepackage[T1]{fontenc}

\DeclareRobustCommand{\VAN}[3]{#2}
\let\VANthebibliography\thebibliography
\def\thebibliography{\DeclareRobustCommand{\VAN}[3]{##3}\VANthebibliography}

\usepackage{graphicx}	% Including figure files
\usepackage{amsmath}	% Advanced maths commands
\title[Simulations of super-critical accretion]{Long timescale numerical simulations of large, super-critical accretion discs}

\author[P. C. Fragile et al.]{
P. Chris Fragile,$^{1,2}$\thanks{E-mail: fragilep@cofc.edu (PCF)}
Matthew J. Middleton,$^{2}$
Deepika A. Bollimpalli$^{3}$
and Zach Smith$^{1}$
\\
$^{1}$Department of Physics and Astronomy, College of Charleston, 66 George Street, Charleston, SC 29424, USA\\
$^{2}$Department of Physics and Astronomy, University of Southampton, Highfield, Southampton SO17 1BJ, UK\\
$^{3}$Department of Astronomy, Astrophysics \& Space Engineering, Indian Institute of Technology Indore, Simrol, Indore 453552, Madhya Pradesh, India
}

\date{Accepted XXX. Received YYY; in original form ZZZ}

\pubyear{2025}

\begin{document}
\label{firstpage}
\pagerange{\pageref{firstpage}--\pageref{lastpage}}
\maketitle

\begin{abstract}
In this paper, we report on three of the largest (in terms of simulation domain size) and longest (in terms of duration) 3D general relativistic radiation magnetohydrodynamic simulations of super-critical accretion onto black holes. The simulations are all set for a rapidly rotating ($a_* = 0.9$), stellar-mass ($M_\mathrm{BH} = 6.62 M_\odot$) black hole. The simulations vary in their initial target mass accretion rates (assumed measured at large radius), with values sampled in the range $\dot{m}=\dot{M}/\dot{M}_\mathrm{Edd} = 1$-10. We find in practice, though, that all of our simulations settle close to a net accretion rate of $\dot{m}_\mathrm{net} = \dot{m}_\mathrm{in}-\dot{m}_\mathrm{out} \approx 1$ (over the radii where our simulations have reached equilibrium), even though the inward mass flux (measured at large radii) $\dot{m}_\mathrm{in}$ can exceed 1,000 in some cases. This is possible because the outflowing mass flux $\dot{m}_\mathrm{out}$ adjusts itself to very nearly cancel out $\dot{m}_\mathrm{in}$, so that at all radii $\dot{M}_\mathrm{net} \approx \dot{M}_\mathrm{Edd}$. In other words, these simulated discs obey the Eddington limit. We compare our results with the predictions of the slim disc (advection-dominated) and critical disc (wind/outflow-dominated) models, finding that they agree quite well with the critical disc model both qualitatively and quantitatively. We also speculate as to why our results appear to contradict most previous numerical studies of super-critical accretion.
\end{abstract}


%%%%%%%%%%%%%%%%%%%%%%%%%%%%%%%%%%%%%%%%%%%%%%%%%%

%%%%%%%%%%%%%%%%% BODY OF PAPER %%%%%%%%%%%%%%%%%%

\section{Introduction}
\label{sec:introduction}

Super-critical accretion, where mass is fed into a system above the nominal Eddington limit, plays a crucial role in many astrophysical settings. It may be a factor in the formation of the first supermassive black holes \citep{Volonteri05, Schneider23, Bennett24}; it likely governs the early evolution of tidal disruption events \citep[TDEs;][]{Dai18, Wu18}; and it is important for understanding the observational appearance of ultra-luminous X-ray sources \citep[ULXs;][]{King01, Kaaret17, King23}. 

The Eddington limit is defined as that state in which there is a perfect balance between the gravitational force attracting matter to a central object and the outward radiation force coming from that object. Assuming an electron-scattering opacity, a pure hydrogen composition, and spherical symmetry gives the standard expression 
\begin{equation}
L_\mathrm{Edd} = \frac{4\pi GM_\mathrm{BH} m_\mathrm{p} c}{\sigma_T} = 1.3 \times 10^{38} \left(\frac{M_\mathrm{BH}}{M_\odot}\right) ~\mathrm{erg~s}^{-1} ~.
\label{eqn:Ledd}
\end{equation}
If we assume the luminosity is powered by accretion onto a black hole, then it is common to take $L_\mathrm{Edd} = \eta \dot{M}_\mathrm{Edd} c^2$, where $\dot{M}_\mathrm{Edd}$ is the corresponding Eddington mass accretion rate and $\eta$ is the radiative efficiency of the disc.

Super-critical accretion has been widely studied, both from an observational perspective and theoretically \citep[see][and references therein]{Kaaret17, King23}. The fundamental issue with super-critical accretion is that, if all the gravitational binding energy of the accreting matter were liberated locally in the form of radiation, as in the standard disc model, then the radiative forces would exceed the gravitational ones, and the disc cannot be in balance. Broadly speaking, two classes of solutions have been proposed to address this problem. The first posits that not all of the energy is actually radiated locally; instead, some of it is advected into the black hole before it has time to escape\footnote{Note that for accreting objects that have physical surfaces, such as neutron stars, all of the liberated accretion energy must ultimately escape.}. The most popular solution within this class is the so-called ``slim'' disc \citep{Abramowicz88, Beloborodov98, Sadowski09}. In the second class of solutions, the excess liberated energy is used to drive an outflow, effectively limiting the amount of matter that actually accretes to smaller radii \citep{Shakura73}. One example of a solution in this class is the ``critical'' disc \citep{Fukue04}. There are also models that combine some degree of advection and outflow \citep{Fukue04, Poutanen07}.

Super-critical accretion has also been studied numerically \citep[e.g.,][]{Ohsuga05, Jiang14, Sadowski16, Takahashi18, Asahina22, Utsumi22}. However, all previous numerical studies differ from the work we present in one or more crucial aspects. For instance, many studies were performed in two-dimensions using an explicit viscosity \citep[e.g.,][]{Ohsuga05, Kitaki21, Hu22, Yoshioka22}. Such simulations preclude any magnetohydrodynamic (MHD) turbulence, magnetically driven outflows, and any non-axisymmetric effects. Others were initialized with a finite torus of gas \citep[e.g.,][]{Jiang14, Sadowski16, Utsumi22}. Such simulations can never truly achieve a global steady state, as the mass reservoir is continuously depleted. More importantly, most of these simulations started with tori that were smaller than their corresponding trapping radius $r_\mathrm{tr} \sim \dot{m}_\mathrm{BH} r_g$, where $r_g = GM_\mathrm{BH}/c^2$ is the gravitational radius and $\dot{m}_\mathrm{BH}$ is the mass feeding rate measured at the black hole\footnote{Throughout this paper, $\dot{m}$ refers to mass accretion rates scaled to Eddington, i.e., $\dot{m} = \dot{M}/\dot{M}_\mathrm{Edd}$.}, possibly forcing them to favor the advective, rather than outflow, solution \citep{Kitaki21, Yoshioka22}. Since our simulations correct many of these issues, we feel they offer an important new perspective in the study of super-critical accretion. 

Since our work focuses on large, steady-state accretion discs, the results are probably most applicable to the case of ULXs. TDEs likely have relatively small discs with rapidly varying mass accretion rates, whereas ULXs have comparatively large discs and more stable accretion rates \citep[although see][]{Middleton22}. The numerical simulations reported in this paper have been specifically designed to match the latter conditions.

In this paper, we first describe our numerical setup (Section \ref {sec:setup}), then highlight results regarding the actual feeding rate of material onto the black hole, as well as the luminosity of the disc (Section \ref{sec:results}). We also compare our results to the two broad classes of super-critical accretion models (Section \ref{sec:models}) and compare our results with previous numerical studies (Section \ref{sec:previous}). We end with our concluding thoughts (Section \ref{sec:conclusions}).

\section{Numerical Setup}
\label{sec:setup}

All of our simulations are performed using the general relativistic radiation MHD (GRRMHD) code Cosmos++ \citep{Anninos05}. We use high-resolution shock-capturing \citep{Fragile12} to solve for the flux and gravitational source terms of the gas and radiation; for the magnetic fields, we evolve the magnetic vector potential \citep{Fragile19}; and for the radiation, we use the (gray opacity) $\mathbf{M}_1$ closure scheme \citep{Fragile14}. Together, these allow us to evolve the following 12 conserved fields: the fluid density $D = W\rho$, the fluid total energy density ${\cal E} = - \sqrt{-g} {T}^0_0$, the fluid momentum density ${\cal S}_j = \sqrt{-g} {T}^0_j$, the magnetic vector potential ${\cal A}_i$, the radiation total energy density ${\cal R} = \sqrt{-g}R^0_0$, and the radiation momentum density ${\cal R}_j = \sqrt{-g} R^0_j$, where $W = \sqrt{-g}u^t$ is the generalized boost factor, $g$ is the four-metric determinant, $\rho$ is the rest-mass density, $u^\mu$ is the fluid four-velocity, $T^{\mu\nu}$ is the fluid stress-energy tensor, and $R^{\mu\nu}$ is the radiation stress-energy tensor. The fluid and radiation fields are coupled through the radiation four-force density \begin{eqnarray}
    G^{\mu} & = &-\rho (\kappa^\mathrm{a}_F+\kappa^s)R^{\mu \nu}u_{\nu} \\
    & &- \rho \left\{ \left[\kappa^s+4\kappa^s\left(\frac{T_{\mathrm{gas}}-T_{\mathrm{rad}}}{m_e}\right) +\kappa^\mathrm{a}_F - \kappa^\mathrm{a}_J \right] \right.  \\
    & & \left.\times R^{\alpha \beta}u_\alpha u_\beta +\kappa^\mathrm{a}_\mathrm{P}a_RT^4_\mathrm{gas}\right\} u^\mu ~,
\label{eqn:Gmu}
\end{eqnarray}
where we assume Planck and Rosseland mean opacities $\kappa^\mathrm{a}_\mathrm{P}=2.8\times10^{23}\,T^{-7/2}_\mathrm{K} \rho_{\mathrm{cgs}}$ cm$^2$ g$^{-1}$ and $\kappa^\mathrm{a}_\mathrm{R}=7.6\times10^{21}\,T^{-7/2}_\mathrm{K}\,\rho_{\mathrm{cgs}}$ cm$^2$ g$^{-1}$, respectively, and $\kappa^\mathrm{s} = 0.34$ cm$^2$ g$^{-1}$ for the scattering opacity, appropriate for solar metallicity with mean molecular weight $\mu=0.615$ and a hydrogen-mass fraction of $X=0.7$. We use the 9D numerical inversion scheme with analytic derivatives from \citet{Fragile14} to recover the primitive fluid and radiation fields. The necessary magnetic field components, including the face-centered, conserved fields ${\cal B}^i$ and zone-centered, primitive field $B^i$ are recovered from the updated vector potential \citep{Fragile19}. 

To initialize our simulations, we start from the \citet{Novikov73} generalization of the Shakura-Sunyaev \citep{Shakura73} thin disc.  As we are only considering a limited radial range, we do not require all three regions of the solution.  Instead, we only initialize the so-called ``inner'' (radiation-pressure-dominated) region, which should exist out to $r \gtrsim 100 r_g$ at the accretion rates we are considering. We follow the form of the Novikov-Thorne solutions given in \citet{Abramowicz13}. This simply requires us to choose a mass for the black hole $M_\mathrm{BH}$, a target mass feeding rate $\dot{m}_0$ measured at large radius, and a Shakura-Sunyaev viscosity parameter $\alpha_\mathrm{SS}$ for the disc. We choose $\alpha_\mathrm{SS} = 0.02$ for our initial setup, though it is difficult to specify a priori what value we should use, as there are multiple possible sources of angular momentum transport in our simulations (magnetohydrodynamic turbulence and magnetically driven winds), and we cannot know ahead of time what effective $\alpha$ they will lead to. Ultimately, however, our goal is just to begin the simulations from some reasonable initial conditions that cover a large radial range. As explained later, we then give the discs plenty of time to approach their true solutions.

From the Novikov-Thorne solution, all we actually require are the radial dependencies of the height $H(R)$ and mid-plane density $\rho_0(R)$ of the disc\footnote{We take $r$ as the spherical-polar radius and $R = r \sin \theta$ as the cylindrical one.}. We also include a small radial velocity $V^R(R)$, associated with the slow inward drift of material through the disc \citep{Penna12}. The initial azimuthal velocity is taken to be Keplerian, $V^\phi(R) = \Omega = (M_\mathrm{BH}/R^3)^{1/2} \left[1+a_*(M_\mathrm{BH}/R^3)^{1/2}\right]^{-1}$.

For the vertical profile, we solve for hydrostatic equilibrium assuming a polytropic EOS with $\Gamma_\mathrm{NT} = 4/3$. The solution yields
\begin{equation}
\rho(R,z) = \rho_0 \left[1 - \frac{z^2}{2H^2}\right]^{1/(\Gamma_\mathrm{NT}-1)}
\end{equation}
and
\begin{equation}
P_\mathrm{tot}(R,z) = \kappa \rho^{\Gamma_\mathrm{NT}} ~,
\end{equation}
where
\begin{equation}
\kappa = \frac{GM_\mathrm{BH} H^2}{\Gamma_\mathrm{NT} (\Gamma_\mathrm{NT}-1) \rho_0^{\Gamma_\mathrm{NT}-1} R^3} ~.
\end{equation}
For the background, we initialize a cold ($e = 3\times 10^{-6} e_\mathrm{max} r^{-2}$), low density ($\rho = 10^{-4} \rho_\mathrm{max} r^{-3/2}$), free-falling ($u^r = -\sqrt{r_\mathrm{BH}/r}$) fluid, where $r_\mathrm{BH} = \left(1 + \sqrt{1 - a_*^2}\right)r_g$ is the radius of the black hole and $a_*$ is its dimensionless spin.

Assuming the gas and radiation are in local thermodynamic equilibrium inside the disc for the initial, analytic solution, we partition the pressure as
\begin{equation}
P_\mathrm{tot} = P_\mathrm{gas} + P_\mathrm{rad} = \frac{k_\mathrm{b}\rho T_\mathrm{gas}}{\bar{m}} + \frac{1}{3}a_\mathrm{R} T_\mathrm{gas}^4 ~,
\end{equation}
where $\bar{m} = \mu m_H$ and $a_\mathrm{R} = 4\sigma_B/c$ is the radiation constant.  We can now solve this quartic equation for $T_\mathrm{gas}(R,z)$. This temperature is also used to set the initial radiation field. In the frame of the fluid, the radiation energy density is taken to be
\begin{equation}
E_\mathrm{rad} = a_R T_\mathrm{gas}^4 ~,
\end{equation}
while the flux, $F^i$, is initially set equal to the gradient of this quantity.  To get the radiation density in the radiation rest frame, $E_R$, and the radiation rest-frame four-velocity, $u_R^\mu$, we follow the transformation procedure outlined in \citet{Sadowski13}. 

One issue with the inner region of the Shakura-Suynaev thin-disc solution is that it is thermally unstable \citep{Shakura76}, as confirmed in earlier numerical work \citep{Jiang13, Mishra16, Fragile18}. One mechanism which can stabilize such discs is the introduction of strong (particularly, toroidal) magnetic fields \citep{Begelman07}, which require particular global magnetic field topologies to maintain \citep{Sadowski16b, Mishra22}. The present simulations start from one such configuration: a zero-net-flux quadrupole field that has two poloidal field loops of opposite polarity stacked vertically, one on top of the other, about the midplane of the disc. The two poloidal loops are greatly elongated in the radial direction, extending from near the inner radius of the disc to nearly the outer boundary of our simulation domain. To initialize this field, we first set the azimuthal component of the vector potential to 
\begin{equation}
A_\phi \propto R^{1.5}z\frac{\sqrt{e^{(-2.5z^2/H^2)}}\sin\left(\pi R/r_\mathrm{max}\right)}{1+e^\Delta} ~,
\label{eqn:potential}
\end{equation}
where $r_\mathrm{max}$ is the maximum radius of the grid, and 
\begin{equation}
\Delta=10\left(\frac{z^2}{H^2}+\frac{(R-R_t)^2}{H^2}-1\right) ~,
\label{eqn:delta}
\end{equation}
where $R_t=\mathrm{max}(r_\mathrm{ISCO},R)$, and $r_\mathrm{ISCO}$ is the usual ISCO\footnote{Innermost Stable Circular Orbit} radius. We then set the poloidal components of the magnetic field as $\mathcal{B}^r = -\partial_\theta A_\phi$ and $\mathcal{B}^\theta = \partial_r A_\phi$. These choices keep the initial magnetic field divergence-free and confined within the initial disc. This particular field configuration is subject to a strong radial shear amplification (leading to a growth of the $\mathcal{B}^\phi$ component) due to the orbital motion of the disc (the so-called $\Omega$-dynamo). Along with the normal, magnetorotational-instability(MRI)-driven amplification, this has been shown to help stabilize similar discs against thermal instability \citep{Sadowski16b, Mishra22}.

The simulations are run on a nested (statically refined), spherical-polar grid with resolution concentrated near the black hole and toward the midplane. We use a logarithmic radial coordinate, $x_1 = 1 + \ln(r/r_\mathrm{BH})$, to cover the range from $0.9\,r_\mathrm{BH} \le r \lesssim 1000\,r_g$. As such, these are the largest three-dimensional, super-critical accretion simulations in terms of the size of the disc that we are aware of, comparable to earlier large-domain, two-dimensional simulations \citep{Kitaki21, Yoshioka22}. The advantage of using such large discs and starting from a Shakura-Sunyaev solution instead of a finite torus is that the simulations can be run for very long times with nearly steady mass accretion rates. It also gives us an opportunity to capture the critical radius, given analytically by \citep{Fukue04, Poutanen07}:
\begin{equation}
r_\mathrm{cr} \approx \frac{5}{3} \dot{m}_0~,
\end{equation}
on the grid, which is the radius inside of which the radiative forces overcome gravity and the traditional disc solution is no longer valid. This has not been the case in most previous numerical work \citep[see][]{Kitaki21}. We include the full polar ($0 \le \theta \le \pi$) and azimuthal ($0 \le \phi \le 2\pi$) domains.%, as we later plan to introduce tilt to these simulations, which means no symmetries can be assumed. 
To improve the resolution near the midplane, a concentrated polar coordinate, $\theta = x_2 + h \sin(2 x_2)$ is used. The base mesh has a resolution of $48\times32\times24$ zones in $\{x_1, x_2, \phi\}$. Outflow boundary conditions are applied at the inner and outer radial limits of the domain, while transmissive boundaries are applied at the poles and periodic boundaries are used in $\phi$. 

\begin{figure*}
\centering
\includegraphics[width=0.65\textwidth]{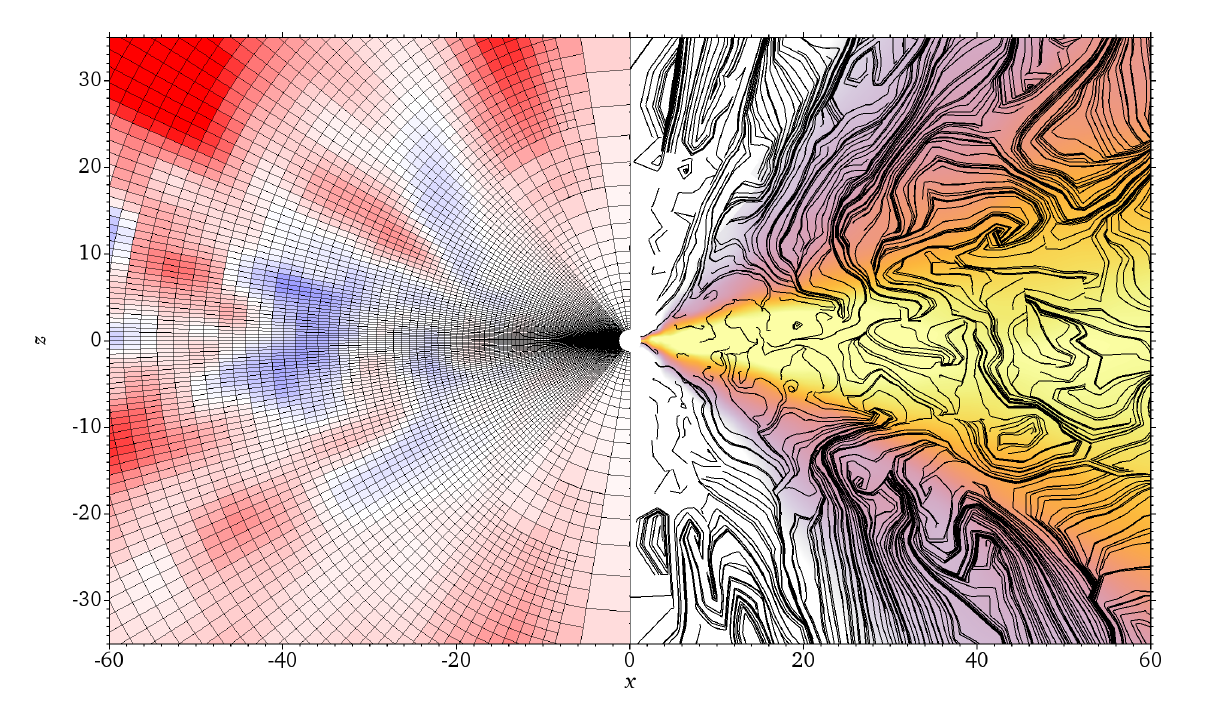}
\caption{Disc and grid configuration at the start of a9r20L3 (the high-resolution interval for simulation a9r20). The {\em left} panel shows the statically refined grid, as well as the radiative flux (arbitrary units). Red colors indicate outgoing flux, while blue colors indicate flux moving toward the black hole. The {\em right} panel shows the logarithm of the gas density, covering 3 orders of magnitude, as well as magnetic field streamlines launched from a uniform sample of points in the plane of this slice.}
\label{fig:initial}
\end{figure*}

As mentioned before, we already know that the Shakura-Sunyaev disc solution is invalid once the mass accretion rate exceeds Eddington, so another option would have been to start our simulations with one of the super-critical disc solutions proposed in Section \ref{sec:introduction}. However, since one of our goals is to assess which class of super-critical solution is applicable to large, steady-state discs, we choose, instead, to start from the Shakura-Suynaev solution and simply give our discs sufficient time to find their preferred super-critical states. To do this in a computationally efficient way, we start all of our simulations on a very low resolution, two-level mesh (base mesh plus one refinement layer for an effective resolution of $96\times64\times48$) and allow them to run to $t_\mathrm{stop} \gtrsim 70,000\,t_g$, where $t_g = GM/c^3$. This is longer than the thermal timescale of the disc ($t_\mathrm{th} \sim [\alpha \Omega]^{-1}$) out to $r \gtrsim 150\,r_g$ and the accretion timescale ($t_\mathrm{acc} \sim r/\vert V^r\vert$) out to $r \gtrsim 20\,r_g$. After this initial ``burn-in'' period, we increase the resolution in one of our simulations by adding another refinement layer before running it for an additional $15,000\,t_g$. A plot of this high-resolution disc and grid is shown in Figure \ref{fig:initial}.

The modest resolutions of our two-level meshes mean that we are not formally resolving the MRI (quality factors $Q_i = \lambda_{\mathrm{MRI},i}/\Delta x_i$ of $Q_\theta \approx 1$ and $Q_\phi \approx 4$, respectively, where $\lambda_{\mathrm{MRI},i} = 2\pi v_{A,i}/|V^\phi|$ is the wavelength of fastest growing MRI mode, $\Delta x_i$ is a typical zone length, and $v_{A,i}=\sqrt{b^ib_i/\rho}$ is Alfv\'en speed in directions $i=\{\theta,\phi\}$). This may lead to our relatively low values for $\alpha\equiv \langle W_{\hat{r}\hat{\phi}}/P_\mathrm{tot}\rangle_t$ of $10^{-3}-10^{-2}$. However, one has to be careful here.  First, we are not using the typical setup of a dipole magnetic field inside a finite torus that has been carefully studied and from which the ``standard'' Q values are mostly derived \citep{Hawley11, Hawley13}. In fact, for our configuration, with a vertically stacked quadrupole field, there is very little $\mathcal{B}^\theta$ to be found in the bulk of the disk. This means our simulations are probably less reliant on the typical axisymmetric modes of the MRI and more dependent on the non-axisymmetric ones, which have been far less studied in terms of saturation values and resolution requirements. Furthermore, with regard to $Q_\phi$, there are two current sheets that form in our simulations, one a little above the midplane and another a little below; this is in contrast to the single current sheet associated with the standard dipole field. This means there will also be regions with relatively weak $\mathcal{B}^\phi$ fields. Finally, since much of the angular momentum transport in these discs is likely in the form of winds, it is unclear how critical the MRI actually is. Additional work will be needed to clarify all of these issues.

In this paper, we report three simulations that vary in their nominal, or intended, mass accretion rate $\dot{m}_0 = \dot{M}/\dot{M}_\mathrm{Edd}$ (assumed measured at large radius), their maximum radial extent $r_\mathrm{max}$, and angular concentration parameter $h$, as detailed in Table \ref{tab:models}. In all other respects, the simulations are the same, with $M_\mathrm{BH}=6.62M_{\odot}$ and $a_* = 0.9$ ($\eta_\mathrm{NT} = 0.156$).

\begin{table*}
\centering
\begin{tabular}{ccccccccccc}
\hline
\hline
& $\dot{m}_0$ & $r_\mathrm{max}/r_g$ & $h$ & $t_\mathrm{stop}/t_g$ & $r_\mathrm{eq}/r_g$ & $\langle\dot{m}_\mathrm{in}(r_\mathrm{eq})\rangle_t$ & $\langle\dot{m}_\mathrm{BH}\rangle_t$ & $\frac{\langle L_\mathrm{out}(r_\mathrm{eq})\rangle_t}{L_\mathrm{Edd}}$ & $\frac{\langle L_\mathrm{kin}(r_\mathrm{eq})\rangle_t}{L_\mathrm{Edd}}$ & $\langle\eta\rangle_t$ \\
\hline
%a9r2 & 0.4 & & & & & & \\
a9r5 & 1 & 300 & 0.12 & 71,625 & 32 & 67 & 1.8 & $\le 6.3$ & $\le 6.4$ & $\le 0.5$ \\
a9r20 & 4 & 1,000 & 0.35 & 165,771 & 49 & 42 & 1.2 & $\le 5.0$ & $\le 2.6$ & $\le 0.7$ \\
a9r50 & 10 & 1,000 & 0.35 & 100,000 & 32 & 23 & 1.9 & $\le 3.8$ & $\le 1.6$ & $\le 0.4$ \\
\hline
\hline
\end{tabular}
\caption{Simulation models and parameters}
\label{tab:models}
\end{table*}

\section{Results}
\label{sec:results}

\subsection{Mass Accretion}

\subsubsection{Black Hole Growth Rates}

If the mass accretion rate $\dot{M}$ is assumed constant throughout, such that the mass accretion rate onto the black hole equals whatever value is fed in at the outer edge of the disc $\dot{M}_0$, then the black-hole mass will grow linearly as
\begin{equation}
M_\mathrm{BH}(t) = M_\mathrm{BH}(t_0) + \dot{M}_0t ~,
\end{equation}
where $M_\mathrm{BH}(t_0)$ is the initial mass, and the growth time will be $\tau_\mathrm{grow} = M_\mathrm{BH}(t_0)/\dot{M}_0$. However, whenever the mass-accretion rate at the outer edge exceeds the Eddington rate, accretion at the inner edge is expected to be suppressed to 
\begin{equation}
\dot{M}_\mathrm{BH} \approx \dot{M}_\mathrm{Edd} ~,
\end{equation}
and the black hole mass grows exponentially as 
\begin{equation}
M_\mathrm{BH}(t) = M_\mathrm{BH}(t_0) e^{t/\tau_\mathrm{grow}} ~,
\end{equation}
where the growth time is now $\tau_\mathrm{grow} \approx M_\mathrm{BH}(t_0)/\dot{M}_\mathrm{Edd}$. Numerically, this corresponds to $\tau_\mathrm{grow} \approx 4.4 \times 10^8 \eta$ yr in the super-Eddington case, which leads to difficulties when trying to understand how black holes can reach masses of up to $10^9 M_\odot$ by the time the universe was $<700$ Myr old \citep{Banados18, Yang21}. So, our first goal with our super-critical simulations is to confirm whether the mass accretion rate onto the black hole really is limited.

\begin{figure}
\centering
\includegraphics[width=1.0\linewidth,trim=0mm 0mm 0mm 0,clip]{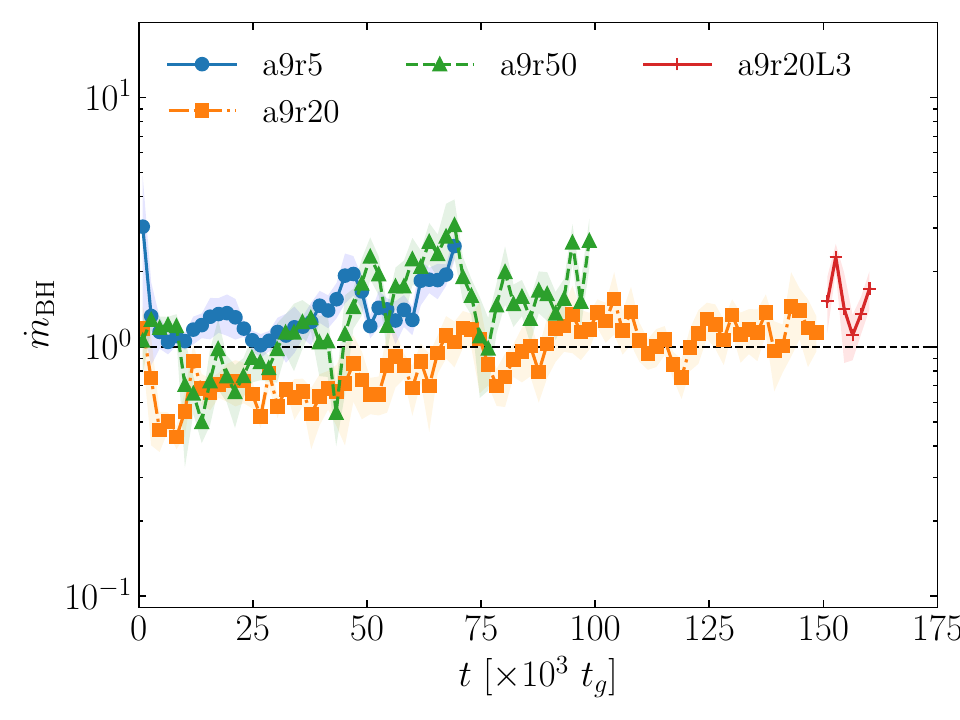}
\caption{Mass accretion rate through the black hole event horizon in units of the Eddington accretion rate $\dot{m}_\mathrm{BH} = \dot{M}_\mathrm{BH}/\dot{M}_\mathrm{Edd}$, smoothed using moving averages over 20 consecutive dumps ($\approx 1,850\,t_g$ in time). The shaded regions show the $1\sigma$ standard deviations, and the black dashed line shows the Eddington limit.}
\label{fig:mdot}
\end{figure}

In Figure \ref{fig:mdot}, we report the time history of mass accretion onto the black hole 
\begin{equation}
\dot{M}_\mathrm{BH}(r_\mathrm{BH},t) = -\int \int \sqrt{-g} \rho u^r {\rm d}\theta {\rm d} \phi
\end{equation}
for all three simulations. The remarkable finding is that they all produce mass accretion rates onto the black hole within a factor of 3 of $\dot{M}_\mathrm{Edd}$ despite covering a full order of magnitude difference in their target value $\dot{m}_0$. The $\dot{m}_\mathrm{BH}$ values are also remarkably steady over time, though there is some evidence for slow secular trends toward increasing $\dot{m}_\mathrm{BH}$ lasting at least $70,000\,t_g$ in all three cases. Additionally, there is maybe a slight jump up in $\dot{m}_\mathrm{BH}$ whenever we increase the resolution of our a9r20 simulation\footnote{Throughout this paper, we refer to the high-resolution extension of simulation a9r20 as a9r20L3.}. Still, the clustering of our results around $\dot{m}_\mathrm{BH} \approx 1$ is noteworthy. 

\subsubsection{How the Eddington limit is achieved}

\begin{figure}
\centering
\includegraphics[width=1.0\linewidth,trim=0mm 0mm 0mm 0,clip]{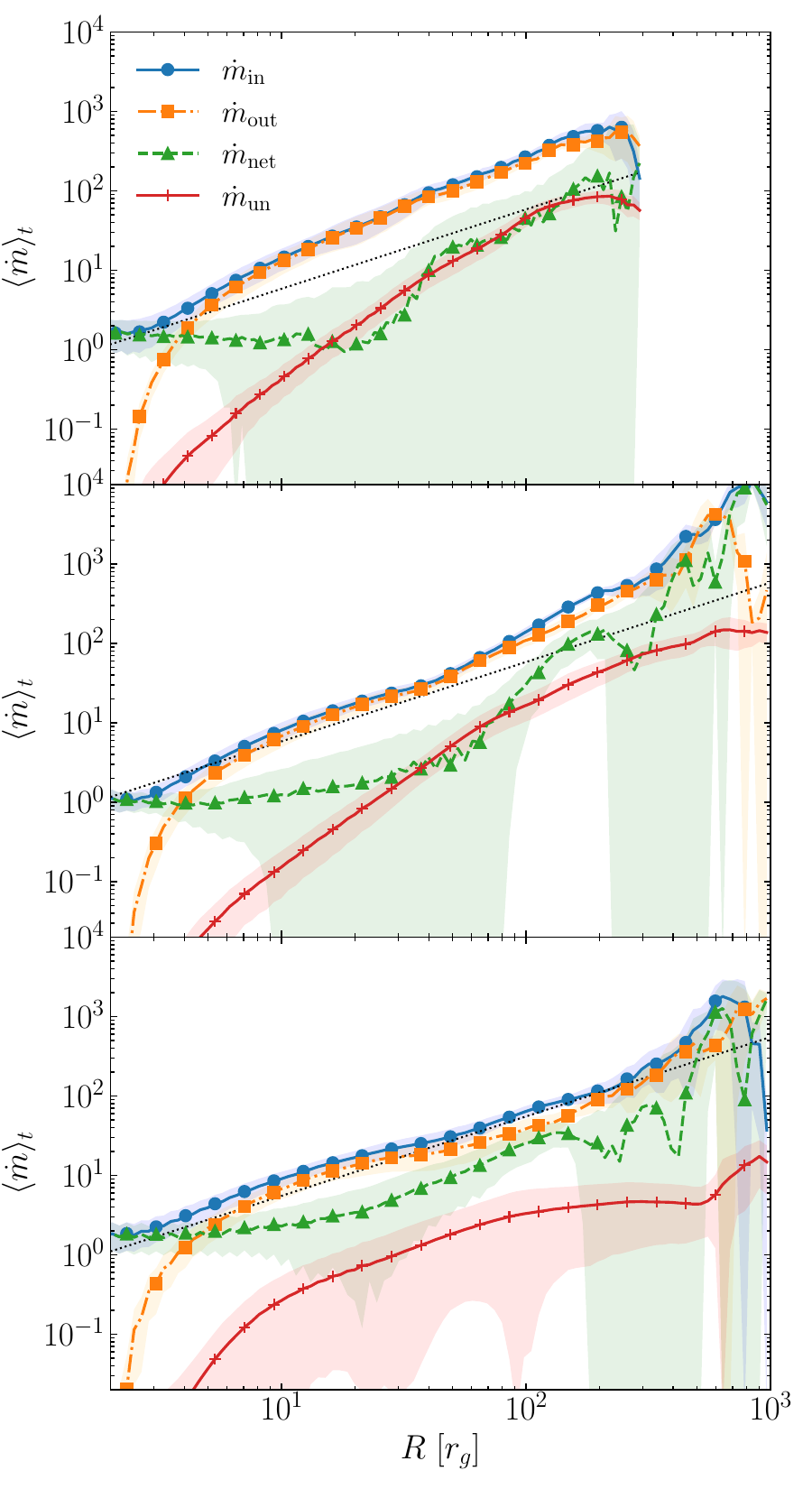}
\caption{Mass fluxes, both inward ($\dot{m}_\mathrm{in}$) and outward ($\dot{m}_\mathrm{out}$), as well as the net mass flux $\dot{m}_\mathrm{net} = \dot{m}_\mathrm{in} - \dot{m}_\mathrm{out}$, all scaled to Eddington and time averaged from $t = 50,000\,t_g$ or $100,000\,t_g$ to $t_\mathrm{stop}$ for the a9r5 ({\em top}), a9r20 ({\em middle}), and a9r50 ({\em bottom}) simulations. The other curves report that portion of $\dot{m}_\mathrm{out}$ that has a positive Bernoulli parameter ($\dot{m}_\mathrm{un}$) and an analytic estimate for $\dot{m}_\mathrm{in}(r) = [\dot{m}_\mathrm{in}(r_\mathrm{cr})-\dot{m}_\mathrm{BH}]r/r_\mathrm{cr}$ (black, dotted curve). The shaded regions show $1\sigma$ standard deviations.}
\label{fig:mdot_flux}
\end{figure}

It is very instructive to see how each of these simulations achieves these nearly identical values of $\dot{m}_\mathrm{BH}$. Figure \ref{fig:mdot_flux} shows time-averaged radial profiles of mass flux, both inward
\begin{equation}
\dot{M}_\mathrm{in}(r,t) = -\int \int \sqrt{-g} \rho \mathrm{min}\{u^r,0\} {\rm d}\theta {\rm d} \phi
\end{equation}
and outward
\begin{equation}
\dot{M}_\mathrm{out}(r,t) = \int \int \sqrt{-g} \rho \mathrm{max}\{u^r,0\} {\rm d}\theta {\rm d} \phi ~,
\end{equation}
for all three simulations. These data are time averaged from $t=50,000\,t_g$ until $t_\mathrm{stop}$ for simulations a9r5 and a9r50 and from $t=100,000\,t_g$ until $t_\mathrm{stop}$ for simulation a9r20. We also plot $\dot{M}_\mathrm{net} = \dot{M}_\mathrm{in} - \dot{M}_\mathrm{out}$, which is an important quantity, as in a steady state, this should be constant as a function of radius (regardless of what super-critical disc model applies). Thus, we can see from Figure \ref{fig:mdot_flux} that we have achieved a reasonably steady state out to $\gtrsim 30 r_g$ in all three cases. Lastly, Figure \ref{fig:mdot_flux} includes $\dot{M}_\mathrm{un}$, which represents that portion of $\dot{M}_\mathrm{out}$ that has a positive Bernoulli parameter $Be = -(T^t_t + R^t_t + \rho u^t) > 0$ \citep{Sadowski16} and thus is likely to be unbound and ultimately escape to infinity. The fact that $\dot{m}_\mathrm{out}$ significantly exceeds $\dot{m}_\mathrm{un}$ in Figure \ref{fig:mdot_flux} implies that much of the material moving outward on our computational domain may eventually turn around and fall back toward the black hole. However, using the Bernoulli parameter to define the unbound outflow is a fairly conservative estimate, as it is possible for matter to be launched with a negative $Be$, yet receive additional acceleration and ultimately escape \citep{Yoshioka22}. As this does not happen within our computational domain, the ultimate fate of this material remains uncertain. As a final note on how these quantities are measured, we emphasize that the mass outflow rates ($\dot{M}_\mathrm{out}$ and $\dot{M}_\mathrm{un}$) are cumulative; in other words, at any given radius they could include matter launched from that or any interior radius. They simply report how much mass is moving outward through a given radius at a given time, irrespective of where it launched from.

An important takeaway from Figure \ref{fig:mdot_flux} is that $\dot{m}_\mathrm{in}$ and $\dot{m}_\mathrm{net}$ both approach 1 at the inner boundary of the computational domain (i.e., at the black hole event horizon). This is achieved despite the fact that $\dot{m}_\mathrm{in}$ can be quite large (easily $>100$) at large radius. This is possible because the mass outflow $\dot{m}_\mathrm{out}$ carefully balances the inflow (compare the blue and orange curves in each panel). In fact, the magnitudes of $\dot{m}_\mathrm{in}$ and $\dot{m}_\mathrm{out}$ are so large and the balance so finely tuned that the difference between the two, $\dot{m}_\mathrm{net}$, shows large statistical fluctuations, particularly on the low side, since it often changes sign (explaining the large green shaded regions in the top two panels of the figure).

Table \ref{tab:models} reports values for $\langle\dot{m}_\mathrm{BH}\rangle_t$, the time-averaged mass accretion rates onto the black hole for each simulation. These results strongly suggest that $\dot{M}_\mathrm{Edd}$ is a meaningful limit for these simulations, and the discs adjust as necessary to meet it. As mentioned previously, this could have major implications for the growth of super-massive black holes in the very early universe. It is also a somewhat surprising result in that it disagrees with practically all previous numerical simulations of super-critical accretion, a point we return to in Section \ref{sec:previous}.

The reader may wonder how the inward mass accretion rates in Figure \ref{fig:mdot_flux} can exceed our reported values for $\dot{m}_0$ by an order of magnitude or more at large radii. First, it may help to emphasize that each $\dot{m}(r)$ comes from an integral over a full $4\pi$ steradian shell. So, especially in the outer disc where there is a lot of mass, if the disc simply sloshes around, it will appear as very large values of $\dot{m}$ (both inward and outward). Additionally, since none of our simulations have reached steady-state solutions at large radii, what we are seeing there may prove to just be an unfortunate transient state set up by our imperfect initial conditions. However, we find it reassuring that the total mass within our computational domain drops by less than 8\% even in our longest duration simulation, so despite what appear to be very large fluxes, we are not actually gaining or losing that much mass compared to what we start with; it is just moving around a lot.

\subsection{Radiative Luminosity}
\label{sec:luminosity}

By definition, ULXs are suspected to be examples of super-critical accretion. The functional definition of a ULX is an off-nucleus, X-ray point source with a luminosity $L_\mathrm{X} > 10^{39}$ erg s$^{-1}$. This limit is chosen because it lies, more or less, at the Eddington limit for a stellar mass object [compare to eq. (\ref{eqn:Ledd})], meaning that ULXs either represent normally accreting objects with mass above what is expected for a stellar remnant (possibly an intermediate mass black hole), or they are stellar remnants apparently emitting above their Eddington limit. We now know that at least some ULXs host neutron stars \citep[i.e., stellar remnants, e.g.][]{Bachetti14, Furst16} and suspect others host stellar-mass black holes \citep{Middleton13, Middleton17, Cseh14}, so we take ULXs as at least one example of a steadily accreting super-critical system to which our results may apply.

Since the defining characteristic of ULXs is that they have apparent, isotropic X-ray luminosities at or above the Eddington limit, it is important for us in this study to look at what radiative luminosity we get from each of our simulations and how that luminosity is distributed in space \citep[since we do not expect ULXs to appear ultra-luminous from all directions;][]{Begelman06, Middleton21}.

\begin{figure}
\centering
\includegraphics[width=1.0\linewidth,trim=0mm 0mm 0mm 0,clip]{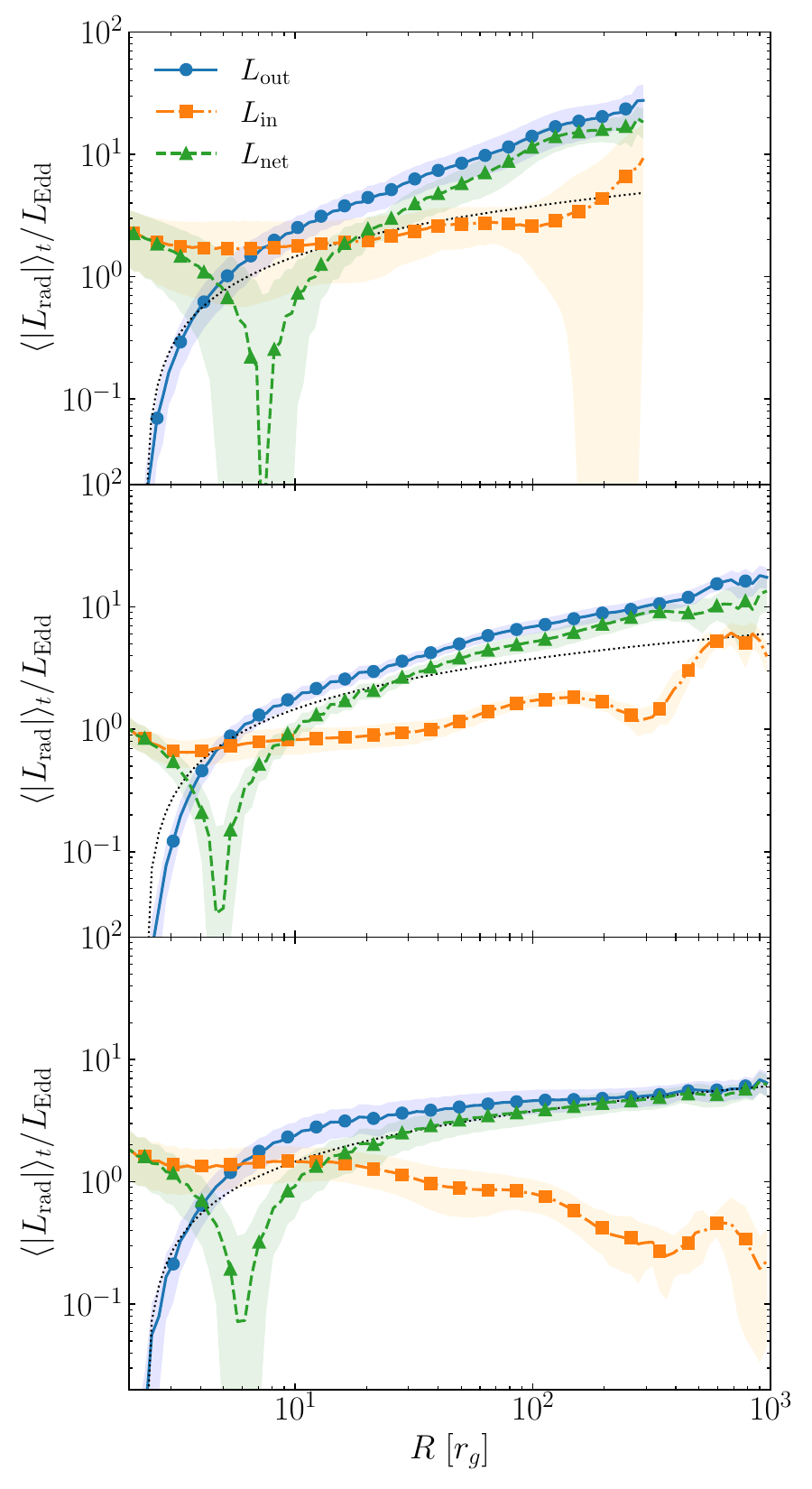}
\caption{Radiative luminosity, both outward ($L_\mathrm{out}$) and inward ($L_\mathrm{in}$), as well as the net luminosity $L_\mathrm{net} = L_\mathrm{out} - L_\mathrm{in}$, all scaled to Eddington and time averaged from $t = 50,000\,t_g$ or $100,000\,t_g$ to $t_\mathrm{stop}$ for the a9r5 ({\em top}), a9r20 ({\em middle}), and a9r50 ({\em bottom}) simulations. The black, dotted curve reports an analytic estimate for $L_\mathrm{out}(r) = \ln(r/r_\mathrm{ISCO})$. The shaded regions show $1\sigma$ standard deviations. The trapping radius $r_\mathrm{tr}$ is apparent as the sharp dip in $L_\mathrm{net}$ around $r \approx 5 r_g$, where it actually changes sign from inflowing (for $r<r_\mathrm{tr}$) to outflowing (for $r>r_\mathrm{tr}$).}
\label{fig:Lrad}
\end{figure}

In Figure \ref{fig:Lrad}, we report the time-averaged radiative luminosity
\begin{equation}
L_\mathrm{rad}(r,t) = -\int \int \sqrt{-g} R^r_t {\rm d}\theta {\rm d} \phi ~,
\end{equation}
integrated over the full $4\pi$ steradians. We report both the outward ($u^r_R > 0$) and inward ($u^r_R < 0$) contributions as a function of radius for all three simulations. The inward luminosity is attributable to photons that are trapped within the accreting gas. The net luminosity, $L_\mathrm{net} = L_\mathrm{out} - L_\mathrm{in}$, reflects the difference between these two components. 

Generally, we find that the overall (outward) radiative luminosity is a few times $L_\mathrm{Edd}$, consistent with expectations for a super-critical accretion disc. However, just as the inward luminosity consists of radiation that is trapped in the accreting gas, some of the outward luminosity may also be trapped in the optically thick wind, some of which is still bound and may fall back to the disc. For this reason, our $L_\mathrm{out}$ likely represents an upper limit of what an observer may measure. Also the luminosities in Figure \ref{fig:Lrad} represent integrals over the complete radial shell, so they are true, total luminosities, and are thus unlikely to match what an observer would infer from any one particular viewing angle.

Another point regarding the radiative luminosity (Figure \ref{fig:Lrad}) is that the net value $L_\mathrm{net}$ changes sign between 5 and $8 r_g$ for all of our simulations, with most of the radiation moving toward the black hole inside that radius and away from the black hole outside it. This dip represents the trapping radius $r_\mathrm{tr}$ for each of our simulations. We note that this trapping radius is relatively close to the inner edge of the disc, so we conclude that advection is not a prominent source of cooling beyond about $20 r_g$ in our simulations. Also, to be clear, there is still some $L_\mathrm{out}$ even inside $r_\mathrm{tr}$, as can be seen in Figure \ref{fig:Lrad}. The point is, though, there is more $L_\mathrm{in}$ than $L_\mathrm{out}$, so in terms of cooling the gas, advection is dominant in that region.

\begin{figure*}
\centering
\includegraphics[width=1.0\textwidth,trim=0mm 0mm 0mm 0,clip]{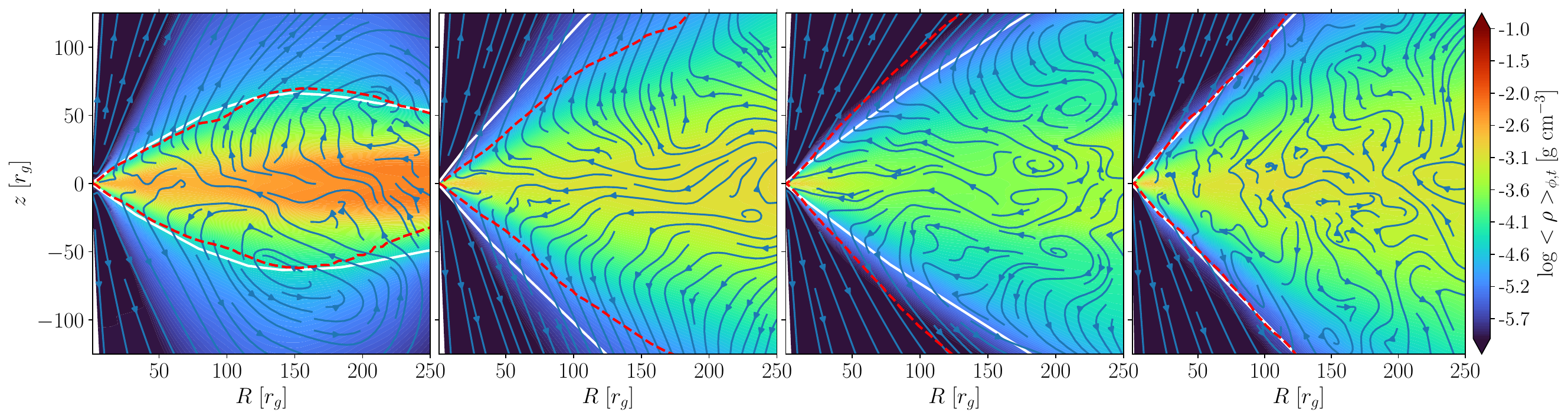}
\caption{Psuedocolor plot of time- and azimuthally averaged gas density and fluid velocity streamlines for simulations a9r5 (first panel), a9r20 (second panel), a9r50 (third panel), plus the high-resolution extension a9r20L3 (last panel). The white lines represent the effective photospheres, while the red, dashed lines delineate the $Be = 0$ boundaries. Time averaging is over the intervals from $t = 50,000\,t_g$, $100,000\,t_g$, or $159,000\,t_g$ to $t_\mathrm{stop}$, depending on the simulation.}
\label{fig:photosphere}
\end{figure*}

An important distinction between optically thick accretion discs and stellar objects is that we do not expect the radiation from discs to be isotropic. Rather, we expect most of it to come out within an optically thin cone centered about the black hole spin axis. Figure \ref{fig:photosphere} shows that the region around the pole in each case is both relatively evacuated of material and lies outside the effective photosphere of the disc, so is optically thin. We locate the effective photosphere by integrating the quantity $-(u_t + u_r)\kappa_\mathrm{e}\rho$ inward from the outer radial boundary of the simulation domain along lines of constant $\theta$ until we reach values $\ge 1$, where the effective opacity is $\kappa_\mathrm{e} = \sqrt{0.5\kappa^\mathrm{a}_\mathrm{R}\kappa^\mathrm{s} }$. 

\begin{figure}
\centering
\includegraphics[width=1.0\linewidth,trim=0mm 0mm 0mm 0,clip]{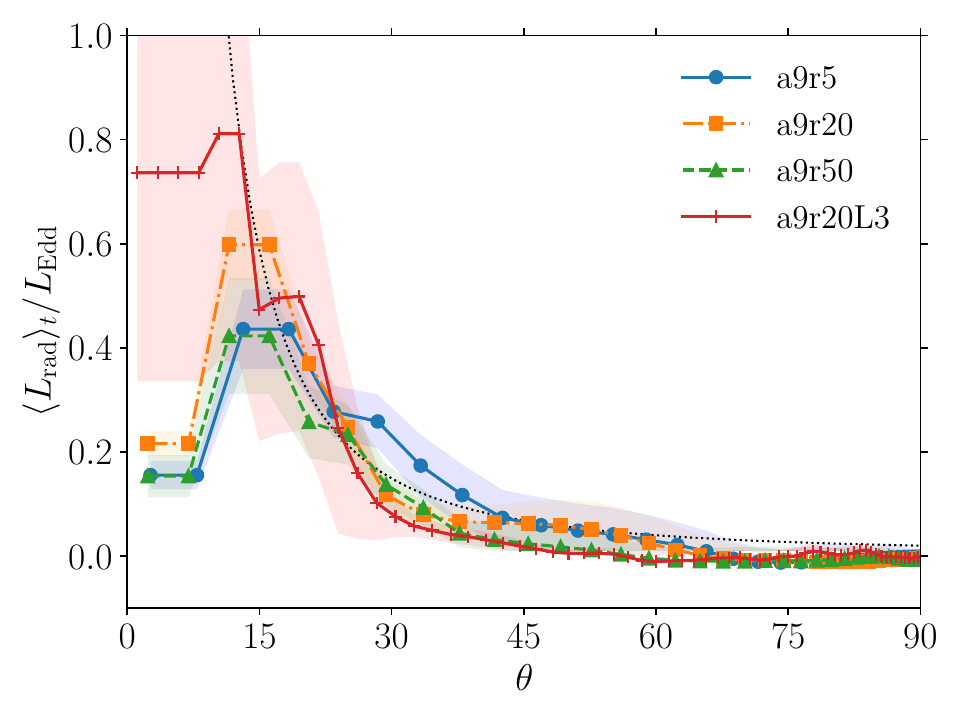}
\caption{Contribution to the radiative luminosity measured at $r_\mathrm{eq}$, broken down into polar angle bins, showing that most of the radiation escapes close to the poles. The black, dotted curve suggests $L_\mathrm{rad}(\theta) \propto 1/(1-\vert\cos\theta\vert)$. Data are time averaged over the intervals from $t = 50,000\,t_g$, $100,000\,t_g$, or $159,000\,t_g$ to $t_\mathrm{stop}$, depending on the simulation, and the shaded regions show $1\sigma$ standard deviations.}
\label{fig:Lrad_theta}
\end{figure}

We can also measure how much radiation is escaping at different angles with respect to the black hole spin axis. We show results for this in Figure \ref{fig:Lrad_theta}. Not surprisingly, near the poles, the luminosity is orders of magnitude greater than in the equatorial plane. This provides a simple explanation for why some suspected ULXs, even within our own galaxy, do not appear to us as such \citep{Begelman06, Middleton21, Veledina24}.  Interestingly, all of our simulations show very similar $\theta$ profiles in Figure \ref{fig:Lrad_theta}, meaning they would all appear to be roughly the same luminosity, when viewed from the same inclination. %perhaps in contrast to \citet{Yoshioka22}, who found their higher $\dot{m}$ simulations to be more beamed.
One odd feature, however, is the drop in the radiative flux right along the pole. We note, though, that similar drops have been seen in other numerical studies \citep[e.g.,][]{Jiang14, Sadowski14, Utsumi22}. In our case, this may have to do with our use of the $\mathbf{M}_1$ closure, although that explanation would not apply to \citet{Jiang14}. Also, the drop is not as pronounced in the high-resolution extension simulation a9r20L3, suggesting this could also be a resolution issue near the pole.

\subsection{Kinetic Luminosity}

\begin{figure}
\centering
\includegraphics[width=1.0\linewidth,trim=0mm 0mm 0mm 0,clip]{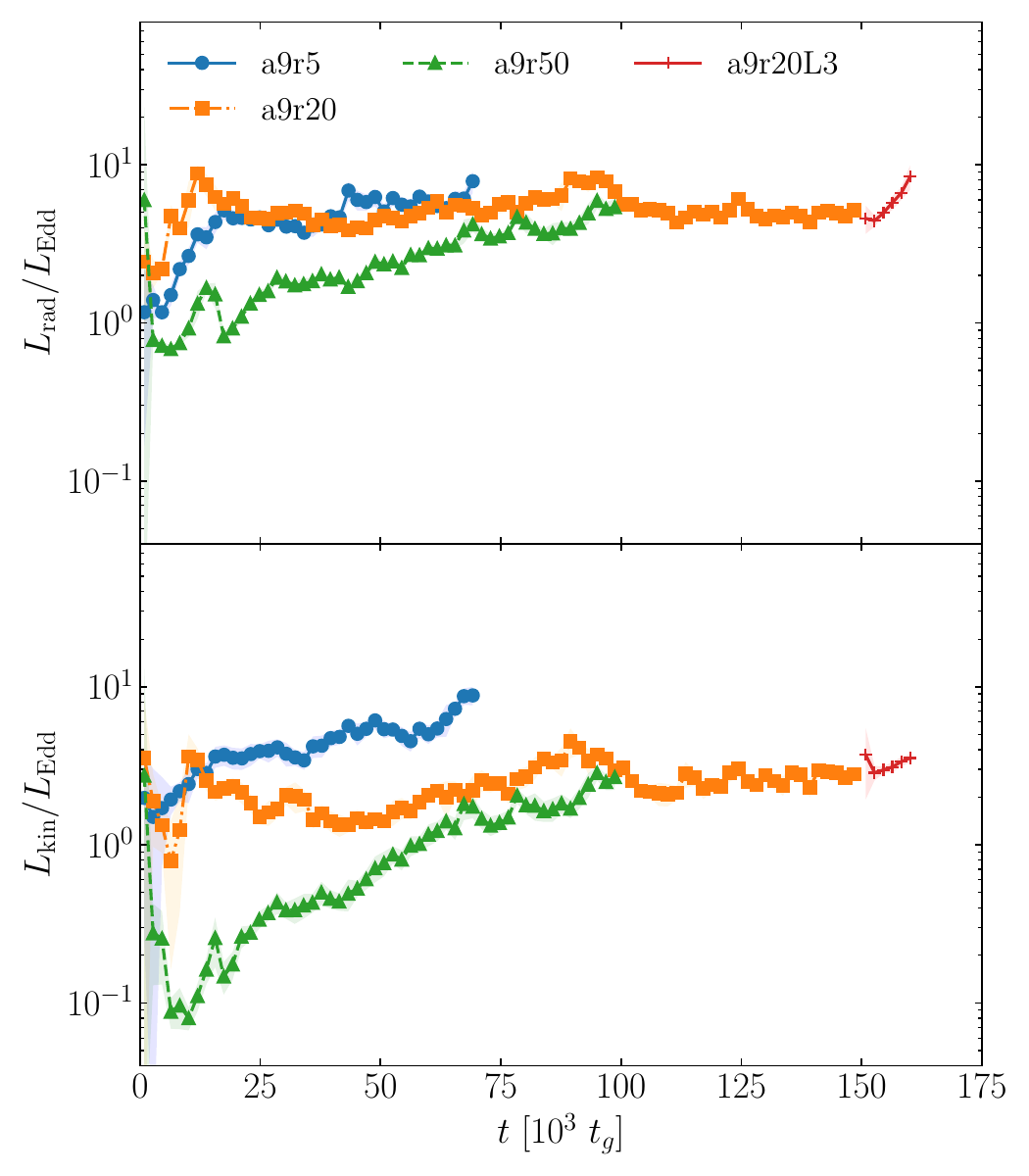}
\caption{Radiative ({\em top panel}) and kinetic ({\em bottom panel}) luminosities as a function of time measured at $r_\mathrm{eq}$ for each simulation. Data have been smoothed by using a moving boxcar averaging window of 20 consecutive dumps. The shaded regions show $1\sigma$ standard deviations.}
\label{fig:Lkin}
\end{figure}

Some ULXs are accompanied by optical nebulae \citep[e.g.][]{Kaaret04} or radio bubbles \citep[e.g.][]{Berghea20} with extents of 10–100 parsecs. These nebulae are thought to be powered by the the ULX itself through some combination of radiation and mass outflow. Indeed, there are now convincing observations of both jets \citep{Middleton13, Cseh14} and winds \citep{Middleton14, Middleton15, Pinto16, Kosec21} from ULXs, with inferred kinetic luminosities on a par with the radiative output. Thus, in addition to radiative luminosities, it is important for us to also examine the kinetic luminosities in the simulations, following:
\begin{equation}
L_\mathrm{kin}(r,t) = -\int \int \sqrt{-g} \rho u^r (u_t + \sqrt{-g_{tt}}) {\rm d}\theta {\rm d} \phi ~.
\end{equation}
%\begin{equation}
%L_\mathrm{kin}(r,t) = \dot{M}_\mathrm{un}(r,t) c^2 ~.
%\end{equation}
%\pcf{Maybe calculate $L_\mathrm{kin}$ as in Sadowski, but only include regions with positive $Be$. This is what Utsumi did.} 
We do this in Figure \ref{fig:Lkin}, where we compare the time histories of the radiative and kinetic luminosities. Each luminosity is measured at the maximum radius for which each simulation has come into inflow equilibrium, $r_\mathrm{eq}$, based on $\dot{m}_\mathrm{net}$ being flat in Figure \ref{fig:mdot_flux}. The values for $r_\mathrm{eq}$, $L_\mathrm{out}(r_\mathrm{eq})$, and $L_\mathrm{kin}(r_\mathrm{eq})$ are reported for each simulation in Table \ref{tab:models}. 

The kinetic luminosities are smaller than the radiative ones by about a factor of two for the a9r20 and a9r50 simulations, but are roughly equal for the a9r5 one. This is consistent with the fact that the a9r5 simulation exhibits mass outflows in Figure \ref{fig:mdot_flux} that are significantly stronger than the other simulations.

As mentioned before, we have measured all of our luminosities through the full $4\pi$ steradians, even though some of the outward radiation may be trapped in the bound outflow and potentially fall back to the black hole at larger radii. Notice that in Figure \ref{fig:photosphere}, the $Be=0$ surface often lies very close to the photosphere, suggesting that most of the radiation passing through the photosphere will escape, while at least some of that within it will remain trapped. Thus, our luminosities likely represent the upper limits of what could be observed. For this reason, some other groups have chosen to report luminosities only from their optically thin or unbound regions. In that case, all of their radiation is likely to reach an observer; however, it probably represents a lower limit on the total luminosity since some of the radiation in the optically thick wind should eventually escape as well. Thus, current simulations can really only bracket what the observed luminosity should be.

\section{Comparison with Super-Critical Disc Models}
\label{sec:models}

As mentioned in Section \ref{sec:introduction}, there are two broad classes of super-critical disc models: those based on advective cooling (e.g., slim discs) and those based on radiatively driven outflows (e.g., critical discs). In this section, we attempt to compare our results with these two classes of models to see if our simulations support either one.

One issue we have to settle before we can make such a comparison is what ``input'' mass accretion rate to consider. All analytic models of accretion are based on the assumption that the input mass accretion rate at large radii is fixed. However, even though our simulations have run for extended periods, they have not reached a steady state all the way to their outer boundaries. Therefore, it would not make sense to use the $\dot{m}$ values there as our input mass accretion rates. Likewise, although we started all of our simulations with a target mass accretion rate in mind based on the Shakura-Sunyaev thin-disc model, this $\dot{m}_0$ was a poor guess at best. We had no way of knowing a priori what the effective viscosity (parameterized by $\alpha$) would be. Not surprisingly, the measured values for $\dot{m}_\mathrm{in}$ are quite different, in general, from our target values and are highly radially dependent. Therefore, for the rest of our analysis, we will use as our input mass accretion rate the value of $\dot{m}_\mathrm{in}$ measured at $r_\mathrm{eq}$, where again $r_\mathrm{eq}$ is the maximum radius out to which the net mass accretion rate has reached a steady value. The measured values for $\langle\dot{m}_\mathrm{in}(r_\mathrm{eq})\rangle_t$ are reported for each simulation in Table \ref{tab:models}.

\subsection{Slim Disc Model}

The slim-disc model \citep{Abramowicz88} assumes that all of the supplied gas reaches the black hole. In other words, the inward mass accretion rate $\dot{M}_\mathrm{in}$ is constant as a function of radius and there are no outflows. This is the first sign that our simulations do not agree with this model, as we see significant mass outflow $\dot{M}_\mathrm{out}$ and a highly radially dependent $\dot{M}_\mathrm{in}$ in Figure \ref{fig:mdot}.

Because all of the gas ultimately reaches the black hole in the slim disc model, it necessarily requires some of the radiation to also be advected into the black hole to prevent the outward radiation pressure from overwhelming the inward gravitational force. The prediction is that the photon trapping radius should scale with the mass accretion rate such that $r_\mathrm{tr} \approx \dot{m}_0 r_\mathrm{ISCO}$. Taking our observed value of $\dot{m}_\mathrm{in}(r_\mathrm{eq})$ as the best measure of $\dot{m}_0$ in our simulations, this would predict a trapping radius of $r_\mathrm{tr} \gtrsim 100 r_g$ for the a9r5 and a9r20 simulations, about twenty times further out than what we actually observe in Figure \ref{fig:Lrad}. This is another sign that our simulations do not agree well with the slim-disc model.

\begin{figure}
\centering
\includegraphics[width=1.0\linewidth,trim=0mm 0mm 0mm 0,clip]{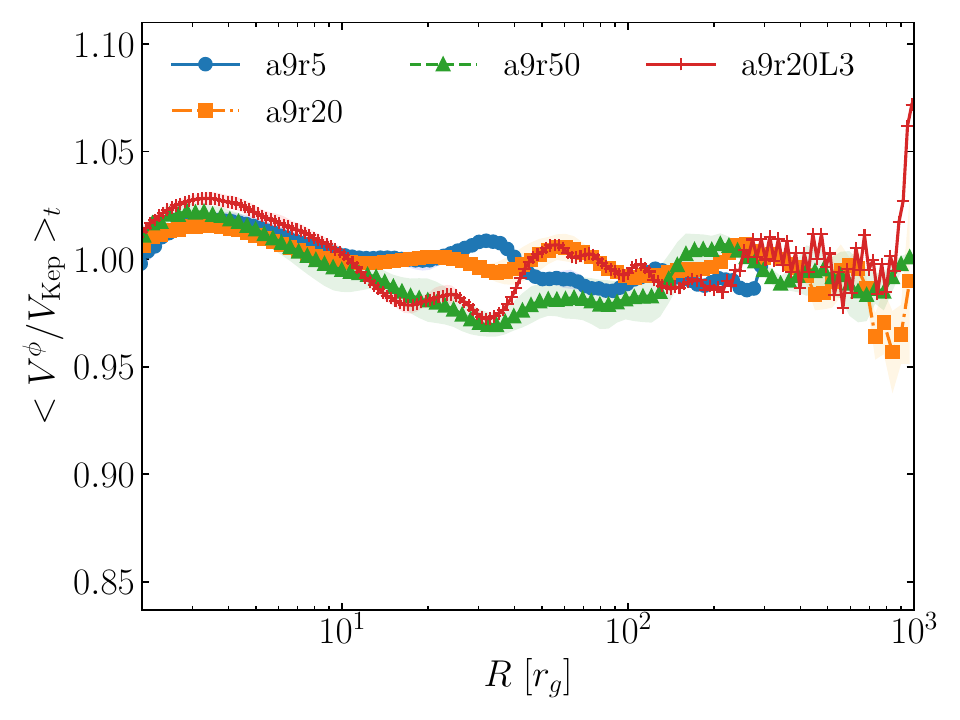}
\caption{Density-weighted, time-averaged angular velocity profiles of the discs divided by a purely Keplerian profile. Our profiles differ by no more than a few percent from purely Keplerian. Data are again time averaged over the intervals from $t = 50,000\,t_g$, $100,000\,t_g$, or $159,000\,t_g$ to $t_\mathrm{stop}$, depending on the simulation, and the shaded regions show $1\sigma$ standard deviations.}
\label{fig:Vphi}
\end{figure}

Another key difference between the slim disc model and the standard Shakura-Sunyaev one is that, while the Shakura-Sunyaev model assumes a purely Keplerian rotation profile, the slim disc requires most of the disc to be slightly sub-Keplerian, with only a small inner, super-Keplerian region \citep{Abramowicz88}. We, instead, find that our discs all have almost perfectly Keplerian rotation profiles (see Figure \ref{fig:Vphi}).

% Also, $H/r$. Our disc is not nearly as thick as often proposed in the literature. $H/r \approx 0.1$ as opposed to 1.

% \begin{figure}
% \centering
% \includegraphics[width=1.0\linewidth,trim=0mm 0mm 0mm 0,clip]{figures/scaleHeight_ave.pdf}
% \caption{Density-weighted, time-averaged scale height profiles of the discs. The black, dotted line shows a constant $H/R$ profile. The shaded regions show $1\sigma$ standard deviations.}
% \label{fig:height}
% \end{figure}

%Somehow need to calculate the contributions of advective cooling.

\subsection{Critical Disc Model}

As mentioned previously, the critical disc model \citep[e.g.,][]{Fukue04} relies on mass outflows to keep the disc below the critical mass accretion rate. Nominally, the outflows should apply to $r < r_\mathrm{cr}$, and it should be the case that $\dot{m}_\mathrm{out}(r < r_\mathrm{cr}) = \dot{m}_\mathrm{in}(r < r_\mathrm{cr}) - \dot{m}_\mathrm{BH}$. In other words, how much matter goes out must match the excess of what is trying to be fed in minus what is actually making it into the black hole. For $r_\mathrm{cr} \gg r_\mathrm{BH}$, this implies $\dot{m}_\mathrm{out}$ should be quite close to $\dot{m}_\mathrm{in}$, which is exactly what we see in Figure \ref{fig:mdot}. In fact, the very large variability in $\dot{m}_\mathrm{net}$ in Figure \ref{fig:mdot} owes to the fact that $\dot{m}_\mathrm{in}$ and $\dot{m}_\mathrm{out}$ have such close numerical values that the difference between them often changes sign. There is also reasonable quantitative agreement between our accretion profiles and the critical disc model, as Figure \ref{fig:mdot_flux} shows that $\dot{m}_\mathrm{in}(r)$ closely follows $[\dot{m}_\mathrm{in}(r_\mathrm{cr})-\dot{m}_\mathrm{BH}]r/r_\mathrm{cr}$ \citep{Poutanen07}.

There are also predictions for how the luminosity should vary for a critical disc inside $r_\mathrm{cr}$. According to \citet{Fukue04}, it should go as $L(r)/L_\mathrm{Edd} \propto \ln(r/r_\mathrm{BH})$, which actually matches the profiles we find for $L_\mathrm{out}$ in Figure \ref{fig:Lrad} fairly well. Another confirmation is the dependence of $L_\mathrm{rad}$ on $\theta$. Figure \ref{fig:Lrad_theta} shows that this agrees with \citep{Fukue11}: $L(\theta)/L_\mathrm{Edd} \propto 1/(1-\vert\cos\theta\vert)$, except right at the poles where the simulation data suddenly drops. Finally, the critical disc model predicts that the disc should maintain a nearly Keplerian velocity profile, consistent with what we report in Figure \ref{fig:Vphi}. To conclude, our $\dot{m}(r)$, $L(r)$, $L(\theta)$, and $V^\phi(r)$ profiles all agree with the predictions of the critical disc model.

\section{Comparison with Other Numerical Work}
\label{sec:previous}

As mentioned in the Introduction, a number of other groups have performed simulations of super-critical accretion discs, and yet, our results appear to be distinct from all previous studies in at least one key aspect: all of our simulations trend towards $\dot{m}_\mathrm{BH} \approx 1$. In other words, our simulations appear to confirm the Eddington limit, whereas other numerical studies do not. In Table \ref{tab:previous}, we provide a sampling of previous simulation results from a variety of other groups \citep[additionally see Table 2 of][]{Toyouchi24}. Noticeable is that all of those simulations found $\dot{m}_\mathrm{BH} \gtrsim 10$.

\begin{table*}
\centering
\begin{tabular}{cccccc}
\hline
\hline
Reference & $\dot{m}_0$ & $\dot{m}_\mathrm{BH}$ & $\eta$ & $L_\mathrm{kin}/L_\mathrm{rad}$ & $r_\mathrm{cr}/r_\mathrm{cir}$\\
\hline
\citet{Jiang14} & $\cdots$ & $\sim22$ & 0.045 & $\sim0.2$ & $\ge 1.5$ \\
\citet{Sadowski16} & $\cdots$ & $\ge 10$ & $\approx 0.03$ & 0-1.4 & $\ge 0.4$ \\
\citet{Abarca18} & $\cdots$ & 22 & $\approx 0.09$ & $\sim 0.1$ & $\ge 0.9$ \\
\citet{Utsumi22} & $\cdots$ & $\gtrsim 10$ & 0.003-0.03 & 0.01-0.4 & $\ge 0.5$ \\
\citet{Yoshioka22} & 35-200 & 11-38 & 0.01-0.02 & 0.02-0.29 & $\gtrsim 0.02$ \\
\hline
\hline
\end{tabular}
\caption{Sampling of published super-critical accretion simulation results. We report the input mass accretion rate $\dot{m}_0$, the measured $\dot{m}_\mathrm{BH}$, the radiative efficiency $\eta$, the ratio of kinetic to radiative luminosities $L_\mathrm{kin}/L_\mathrm{rad}$, and the ratio of the critical radius to the radius of the torus pressure maximum or the circularization radius of the gas $r_\mathrm{cr}/r_\mathrm{cir}$. Since most of these simulations used non- or slowly rotating black holes, we assume a radiative efficiency of 10\% when defining $\dot{M}_\mathrm{Edd}$ in this table. In many cases, we were unable to extract the values of $\dot{m}_0$ from the information provided in the original paper. In those cases, our estimate of $r_\mathrm{cr}$ is based on $\dot{m}_\mathrm{BH}$, which will generally be much smaller than $\dot{m}_0$, making our estimates of $r_\mathrm{cr}$ stringent lower limits.}
\label{tab:previous}
\end{table*}

We have a few ideas about why our simulations may have yielded different results:
\begin{itemize}
\item Most previous simulations started from a finite torus of gas, and in many of them, the critical radius $r_\mathrm{cr}$, where the radiation pressure first exceeds gravity, lies beyond the pressure maximum of the torus. This may prevent the disc from having the necessary space and time to fully adjust to the radiation pressure before accreting. This was already pointed out in \citet{Kitaki21}. 
%2. Maybe it is that $r_\mathrm{cr}$ is larger than $r_\mathrm{eq}$. However, our simulations are not in equilibrium out to $r_\mathrm{cr}$.
%3. Maybe it is that $\dot{m}$ initially rises so rapidly because of how the simulations are constructed that the inner, advection-dominated region effectively shields the outer regions, preventing them from feeling the full effects of the radiation. Probably not.
\item In other cases, it must be that the ratio of the advection timescale to the radiation diffusion timescale is much smaller than in our simulations. This could be due to a loss of angular momentum support, leading to significantly sub-Keplerian angular velocity profiles and short advection times in the other simulations. Or the low $\alpha$ values in our own simulations may lead to unrealistically large advection times. 
\item Another possibility is that the radiation diffusion timescale in the other simulations is much longer, either because of differences between the radiative transport methods or because some of those simulations lack MHD turbulence, which can give the radiation easier channels to escape from the disc.
\item Finally, our unique starting magnetic field topology could also be a contributor. Perhaps some field topologies are more prone to driving Blandford-Payne \citep{Blandford82} winds than others, possibly altering $\dot{M}_\mathrm{out}$, or yield lower saturation values for $\alpha$, altering $\dot{M}_\mathrm{in}$.
\end{itemize}
Since we think our methodology and setup are more appropriate for simulating large, super-critical accretion discs, as may be applicable to ULXs, than any previous simulations, we stand by our finding that such discs are locally Eddington limited at all radii, even when $\dot{m}_0 \gg 1$.

Not surprisingly, since we measure comparable luminosities to previous simulations, but significantly smaller $\dot{m}_\mathrm{BH}$, our discs yield radiative efficiencies that are an order of magnitude or more higher. Using our values for $L_\mathrm{out}(r_\mathrm{eq})$ and $\dot{M}_\mathrm{BH}$, we measure radiative efficiencies of $\langle\eta\rangle_t = 0.3$-0.7. This is somewhat higher than the efficiency expected from thin-disc theory (0.156). However, as mentioned in Section \ref{sec:luminosity}, our values for $L_\mathrm{out}$ should be viewed as upper limits, meaning our values for $\eta$ are also upper limits. To avoid confusion, we remind our readers that our simulations are not done in the magnetically arrested disk (MAD) limit, which can also result in high radiative efficiencies \citep{Thomsen22}.

\section{Conclusions}
\label{sec:conclusions}

We have reported on one of the first sets of large (radially extended), three-dimensional GRRMHD simulations of super-critical accretion onto black holes. This work is most directly applicable to ULX systems, but may also tell us something about the growth history of black holes over cosmic time.

The most notable finding in our work is that all of our simulations trend toward $\dot{m}_\mathrm{BH} \approx 1$. The takeaway is that for super-critical discs fed by thin, Keplerian discs at large radii, it appears $\dot{M}_\mathrm{Edd}$ is a meaningful limit\footnote{The Eddington limit does not apply whenever the angular momentum of the gas is so low that it cannot circularize as a disk before it accretes into the black hole \citep{Fragile12, Inayoshi16}.}, though this should be validated over a wider parameter range. This is in good agreement with long-standing theory, but poses a significant challenge when trying to understand the growth of the first supermassive black holes. Either they cannot grow from steady, long-term accretion from a large, aligned, Keplerian disk or they cannot start from stellar mass accretors.

To help interpret our results, we tested them against two broad classes of models of super-critical accretion: advection-dominated slim discs and outflow-dominated critical discs. We found that our results do not agree with the main predictions of the slim disc, as we see significant mass outflow, a small trapping radius, and nearly perfectly Keplerian velocity profiles. By contrast, our results agree well with the critical disc model, where mass outflow closely balances mass inflow at all radii to produce a net accretion rate close to $\dot{M}_\mathrm{Edd}$. We also found that our luminosity profiles, both in radius $L(r)$ and polar angle $L(\theta)$ match the predictions of the critical disc model.

We caution that it is unclear whether or not we resolve the critical radius $r_\mathrm{cr}$ within our computational domain. One way to identify this radius would be to look for where the profile of $\dot{m}_\mathrm{in}(r)$ (or likewise $\dot{m}_\mathrm{out}(r)$) flattens out (i.e., becomes independent of $r$). Unfortunately, we do not see convincing evidence for such plateaus in Figure \ref{fig:mdot_flux} for any of our simulations. This tells us that $r_\mathrm{cr}$ must lie beyond the equilibrium radius $r_\mathrm{eq}$ achieved in each simulation (see Table \ref{tab:models}). It could be that extending these simulations further in time would allow us to eventually capture $r_\mathrm{cr}$ on the grid, or it could be that we would need to extend the grid even further out in radius. Alternatively, we could try other disc parameters to see if we could bring $r_\mathrm{cr}$ to a smaller radius that is more easily captured. Regardless of how it is accomplished, it is an important goal to try to capture $r_\mathrm{cr}$ within the computational domain, and we will continue to work toward that in future simulations. However, this does not alter any of the conclusions we put forth in this study.

Our result of the Eddington limit being enforced in all our simulations is surprising, as it sits in contrast to nearly all previous numerical simulations of super-critical accretion. We speculated in Section \ref{sec:previous} that this likely has to do with differences in how we set up our simulations compared to all other work. If so, that is an important lesson to consider for anyone thinking of doing simulations of super-critical accretion in the future. One clear point seems to be that if the circularization radius of the gas $r_\mathrm{cir}$ is smaller than the critical radius $r_\mathrm{cr}$, then the disc may not be able to adjust fully to the critical solution and will therefore be forced to favor the advective one, as may be appropriate for TDEs but not ULXs.

\section*{Acknowledgements}

PCF gratefully acknowledges the support of the National Science Foundation through grants PHY-1748958 and AST-1907850 and NASA under award No 80NSSC24K0900. MJM gratefully acknowledges the support of STFC (ST/V001000/1). DAB acknowledges support from IIT-Indore, through a Young Faculty Research Seed Grant (project: `INSIGHT'; IITI/YFRSG/2024-25/Phase-VII/02). This work was performed in part at the Aspen Center for Physics, which is supported by National Science Foundation grant PHY-1607611. This work used the DiRAC Memory Intensive service (Cosma8) at Durham University, managed by the Institute for Computational Cosmology on behalf of the STFC DiRAC HPC Facility (www.dirac.ac.uk). The DiRAC service at Durham was funded by BEIS, UKRI and STFC capital funding, Durham University and STFC operations grants. DiRAC is part of the UKRI Digital Research Infrastructure. 

%%%%%%%%%%%%%%%%%%%%%%%%%%%%%%%%%%%%%%%%%%%%%%%%%%
\section*{Data Availability}

The data underlying this paper will be shared upon reasonable request to the corresponding author.



\bibliographystyle{mnras}
%\bibliography{refs} % if your bibtex file is called example.bib

%%%%%%%%%%%%%%%%%%%%%%%%%%%%%%%%%%%%%%%%%%%%%%%%%%

%%%%%%%%%%%%%%%%% APPENDICES %%%%%%%%%%%%%%%%%%%%%

%\appendix


%If you want to present additional material which would interrupt the flow of the main paper, it can be placed in an Appendix which appears after the list of references.

%%%%%%%%%%%%%%%%%%%%%%%%%%%%%%%%%%%%%%%%%%%%%%%%%%

% Don't change these lines
\bsp	% typesetting comment
\label{lastpage}
\end{document}